\documentclass{PoS}
\usepackage{epsfig}
\usepackage{latexsym}
\usepackage{bm}

\newcommand{\be}{\begin{equation}}
\newcommand{\ee}{\end{equation}}
\newcommand{\ba}{\begin{eqnarray}}
\newcommand{\ea}{\end{eqnarray}}
\newcommand{\bite}{\begin{itemize}}
\newcommand{\eite}{\end{itemize}}

\newcommand{\Dizq}{{\hspace{1mm}}^{{}^{^\leftarrow}}\hspace{-3mm} }

\title{Fermionic correlation functions from the staggered Schr\"odinger
 functional \thanks{IFT-UAM/CSIC-08-68 TCDMATH-08-14}}

\ShortTitle{Fermionic correlation functions from the staggered Schr\"odinger functional }

\author{\speaker{Paula~P\'erez~Rubio}%
\thanks{This research was supported in part by the Universidad Aut\'onoma de Madrid
 and in part by Trinity College Dublin.}\\
        Instituto de F\'isica Te\'orica UAM-CSIC, 28049 Cantoblanco, Spain, \\
	and\\
        School of Mathematics, Trinity College, Dublin 2, Ireland.\\  
        E-mail: \email{%paula.perez@uam.es \\ $\quad$ $\quad $  
	               perez@maths.tcd.ie}}
	
\author{Stefan Sint\\
        School of Mathematics, Trinity College, Dublin 2, Ireland\\
        E-mail: \email{sint@maths.tcd.ie}}

\abstract{
We consider the Schr\"odinger functional with 
staggered one-component fermions on a fine lattice of
size $(L/a)^3 \times (T/a)$ where $T/a$ must be an odd number.
In order to reconstruct the four-component spinors, two different
set-ups are proposed, corresponding to the coarse lattice 
having size $(L/2a)^3 \times (T'/2a)$, with $T' = T \pm a$.
The continuum limit is then defined at fixed $T'/L$. 
Both cases have previously been investigated in
the pure gauge theory. Here we define fermionic correlation functions 
and study their approach to the continuum limit at tree-level 
of perturbation theory.}

\FullConference{The XXVI International Symposium on Lattice Field Theory \\
		 July 14 - 19, 2008\\
		 Williamsburg, Virginia, USA}

\begin{document}

%%%%%%%%%%%%%%%%%%%%%%%%%%%%%%%%%%%%%%%%%%%%%%%%%%%%%%%%%%%%%%%%%%
%%%%%%%%%%%%%%%%%%%%%%%%NEW SECTION%%%%%%%%%%%%%%%%%%%%%%%%%%%%%%%
%%%%%%%%%%%%%%%%%%%%%%%%%%%%%%%%%%%%%%%%%%%%%%%%%%%%%%%%%%%%%%%%%%
\vspace{-5mm}
\section{Introduction}

Establishing a quantitative connection between the low energy and the perturbative regimes of 
QCD is one of the primary tasks for any attempt to solve QCD quantitatively.
For lattice QCD, the main problem consists in the large scale differences involved, which
cannot be resolved on a single finite lattice. A solution to this ``non-perturbative renormalisation
problem" has been proposed a while ago~\cite{Jansen:1995ck} and amounts to apply recursive 
finite size scaling techniques to the renormalised parameters and operators in a suitable renormalisation
scheme. The Schr\"odinger functional (SF)~\cite{Luscher:1992an,Sint:1993un} 
gives rise to a class of such schemes with a number of technical advantages. 
In QCD with zero and two quark flavours, the running 
coupling~\cite{Luscher:1992an,Sint:1995ch,Luscher:1993gh,DellaMorte:2004bc} 
and quark masses\cite{Sint:1998iq,Capitani:1998mq,DellaMorte:2005kg}, as well 
as a range of composite operators have been studied. Note that the final 
results are obtained in the continuum limit and thus independent of the 
details of the lattice regularisation.
Most results have been obtained using the implementation of the SF in 
QCD with Wilson type quarks~\cite{Sint:1993un}. However, for applications 
to QCD with four quark flavours, or for QCD like theories with multiples of four 
fermion flavours staggered fermions appear to be a natural alternative.
Interesting universality tests could be devised and one may expect a better control over the
continuum limit. Hence, in view of applications to four-flavour 
QCD we here revisit the implementation of the Schr\"odinger functional 
for staggered quarks, which has previously been studied 
in~\cite{Miyazaki:1994nu,Heller:1997pn}. Its applications 
have so far been limited to studies of 
the running coupling for QCD-like theories with eight, 
twelve and sixteen fermion flavours~\cite{Heller:1997pn,Appelquist:2007hu}. 
As noticed in~\cite{Miyazaki:1994nu,Heller:1997pn}, the time extent
of the lattice, $T/a$, needs to be odd with staggered quarks, whereas the spatial lattice directions
must have  even extent, $L/a$. This makes it impossible to set $L=T$ exactly, 
and one needs to deal with the resulting O($a$) effects. In order to 
cancel those, Heller~\cite{Heller:1997pn} proposed to average results 
for the gauge coupling obtained in two separate simulations with $T=L\pm a$. 
While this seems to work out for the SF coupling, at least to one-loop 
order in perturbation theory, it is less clear 
how to proceed in the case of fermionic correlation functions.
In particular, one needs to discuss how to reconstruct the four-component spinors in both cases,
only one of which was considered in~\cite{Heller:1997pn}. Ideally, one would like to avoid the
averaging procedure altogether, and it has been shown in the pure gauge theory how this can
be achieved by redefining the approach to the continuum limit~\cite{PerezRubio:2007qz}.

This writeup is organised as follows. We start by reviewing the basics of the 
Schr\"odinger Functional and the definition of fermionic correlation functions. Next we
reconstruct the action in terms of the four-component spinors for both cases, $T'=T\pm a$. A chiral 
rotation is then carried out to recover the standard Schr\"odinger Functional boundary conditions 
for the fermionic fields. We show the results of the computation of the correlation functions
at tree level of perturbation theory and we finish with an outlook to future work.

\section{The Schr\"odinger Functional and correlation functions}

The Schr\"odinger Functional is the Euclidean path integral of QCD on
a hyper cylinder as space-time manifold. 
Dirichlet boundary conditions are imposed at Euclidean times $x_0=0, T$, while all fields are
$L$-periodic in the spatial directions. For the fermionic fields one sets
% The fermionic fields, described classically with a first order 
%differential equation, only have half their degrees of freedom fixed, at each boundary,
%
\ba\label{SFBC}
P_+\psi(y) \big |_{y_0= 0} = \rho({\bf y})&\qquad &P_-\psi(y)\big|_{y_0= T}= \rho'({\bf y}), \nonumber \\
\bar \psi(y) P_-\big |_{y_0 = 0} =\bar \rho({\bf y})&\qquad & \bar \psi(y) P_+\big|_{y_0 = T}= \bar \rho'({\bf y}),
\ea
where $P_\pm = \frac 12 (1 \pm \gamma_0)$. Using a continuum notation, 
the spatial gauge field components satisfy the conditions, 
\be
A_k(y)\big|_{y_0= 0}= C_k \qquad A_k(y)\big|_{y_0 = T} = C_k'.
\ee 
The Schr\"odinger Functional can then be regarded as a functional of the boundary fields,
\be
\mathcal Z [C, C', \rho, \rho', \bar\rho, \bar \rho']= 
\int \mathcal D [A, \psi,\bar \psi] e^{-S[A, \psi, \bar \psi]},
\ee
and expectation values of any product of fields $\mathcal O$, are defined by,
\be
\langle \mathcal O \rangle = \left\{\frac {1}{\mathcal Z}\int D[A, \psi, \bar \psi]
\mathcal O e^{-S[A, \psi, \bar \psi]} \right\}_{\rho= \rho'= 0 ;\, \bar \rho= \bar \rho'= 0}.
\ee
Note that observables may contain quark and antiquark fields at the boundaries,
by including derivatives with respect to the fermionic boundary fields, 
$\zeta({\bf y}) = \frac {\delta }{\delta \bar\rho ({\bf y})}, 
\bar \zeta ({\bf y}) = - \frac{\delta }{\delta \rho({\bf y})},$
and analogously for $\zeta'({\bf y}), \bar \zeta'({\bf y})$. 
Provided the gauge boundary fields are taken to be spatially constant, one
may obtain gauge invariant quark bilinear sources at the boundaries, such as
\be
{\mathcal O}^a = \int {\rm d}^3 {\bf y}'{\rm d}^3{\bf y}^{''}\bar \zeta ({\bf y}')\gamma_5 
 \textrm{$\frac 12$}\tau^a \zeta({\bf y}^{''}),
\ee
where $\tau^a$ is a flavour matrix and both the quark and anti-quark fields 
are projected to zero momentum. Using such sources the simplest 
fermionic correlation functions for the axial vector current and density take
the form~\cite{Luscher:1996sc}
\be
f_{\rm A}^{ab}(y_0)  = - \langle A_0^a(y){\mathcal O}^b \rangle,\qquad 
f_{\rm P}^{ab}(y_0)  = - \langle P^a(y){\mathcal O}^b \rangle,\qquad
f_1^{ab} = - \langle {\mathcal O}^a{\mathcal O}^{'b} \rangle.
\ee
Note that with an exact flavour symmetry all correlation functions would be proportional to $\delta^{ab}$. 
%
%%%%%%%%%%%%%%%%%%%%%%%%%%%%%%%%%%%%%%%%%%%%%%%%%%%%%%%%%%%%%%%%%%
%%%%%%%%%%%%%%%%%%%%%%%%NEW SECTION%%%%%%%%%%%%%%%%%%%%%%%%%%%%%%%
%%%%%%%%%%%%%%%%%%%%%%%%%%%%%%%%%%%%%%%%%%%%%%%%%%%%%%%%%%%%%%%%%%
\vspace{-3mm}
\section{Reconstruction of the four component spinors}
\vspace{-2mm}
\subsection{Case $T' = T-a$}
\begin{figure}[ht!]
\begin{center}
\includegraphics[width=0.3\textwidth]{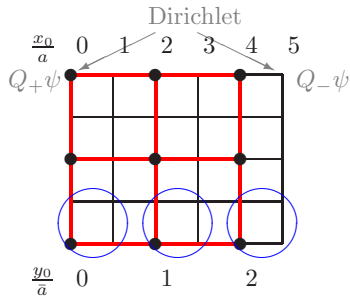}
\vspace{-0.3cm}
\caption{\label{figura1} Reconstruction of the spinors on a $T  = L + a$ lattice.}
\end{center}
\vspace{-0.6cm}
\end{figure}

This case is the one already discussed in\cite{Miyazaki:1994nu,Heller:1997pn}.
The four-component spinors reside in a coarse lattice with lattice spacing $\bar a = 2a$. 
In Figure \ref{figura1}, the thin lines correspond to the fine lattice, and
the dots represent the points of the coarse lattice where the reconstructed fermions live. The 
variable $y$ refers to the points in the coarse lattice, and $x$ to the fine lattice, 
and they are related by $x=2y+a\xi$, with $\xi_\mu$ taking values in $\{0,1\}$. 
%
%
%
%%%%%%%%%%%%%%%%%%%%%%%%%%%%%%%%%%%%%%%%%%%%%%%%%%%%%%%%%%%%%%%%%%%%
%%%%%%%%%%%%%%%%%%%%%%%%%%%%%%FIGURE%%%%%%%%%%%%%%%%%%%%%%%%%%%%%%%%
%%%%%%%%%%%%%%%%%%%%%%%%%%%%%%%%%%%%%%%%%%%%%%%%%%%%%%%%%%%%%%%%%%%%
%%%%%%%%%%%%%%%%%%%%%%%%%%%%%%%%%%%%%%%%%%%%%%%%%%%%%%%%%%%%%%%%%%%%
%%%%%%%%%%%%%%%%%%%%%%%%%%%%%%%%%%%%%%%%%%%%%%%%%%%%%%%%%%%%%%%%%%%%
%%%%%%%%%%%%%%%%%%%%%%%%%%%%%%%%%%%%%%%%%%%%%%%%%%%%%%%%%%%%%%%%%%%%
%
Introducing the transition fields $\chi_\xi(y) = \chi(x)$, $\bar \chi_\xi(y) = \bar \chi (x)$. 
the transformation is specified as
\be
\psi_{\alpha a}(y)= \frac 14 \sum_{\xi}\left( \Gamma_{\xi} \right)_{\alpha a} \chi_{\xi}(y)\qquad
\bar \psi_{a\alpha }(y)= \frac 14 \sum_{\xi}\bar \chi_\xi(y)\left(\Gamma_{\xi} \right)_{a\alpha},
\ee
with $\Gamma_\xi= \frac 12 \gamma_0^{\xi_0}\gamma_1^{\xi_1}\gamma_2^{\xi_2}\gamma_3^{\xi_3}$.
In Figure \ref{figura1}, the one-component fermionic fields which constitute a reconstructed 
quark field are the ones contained in the circles. Note that the Dirichlet boundary conditions 
at $x_0 = 0,T$, imply a projection onto half of the components of the reconstructed quark field.
Labelling the (hermitean) flavour matrices by their $\gamma$-matrix structure, e.g.~$\tau_\mu = \gamma_\mu^T, \tau_{\mu5} = i(\gamma_\mu \gamma_5)^T...$, and denoting the symmetric  derivative by $\tilde \partial_\mu$ and the second derivative by $\Delta_\mu$,  the boundary conditions read
%amount to these 
%boundary conditions for the reconstructed fields, 
%
\ba
Q_+\psi(0,{\bf y}) &=& \hat \rho ({\bf y}), \quad Q_-\psi(T',{\bf y}) = \hat \rho'({\bf y}), \nonumber \\
\bar \psi(0, {\bf y})Q_+ &=& \hat{\bar \rho}({\bf y}), \quad \bar \psi(T',{\bf y})Q_- = \hat{\bar \rho}'({\bf y}),
\ea
with projectors $Q_{\pm} = \frac 12 (1 \pm i\gamma_0\gamma_5\tau_{05})$. For
homogeneous boundary conditions, and with all fields at times $x_0<0$ and $x_0>T'$ set to zero, 
the reconstructed action takes the form, 
\be\label{action1}
S_{SQ}^{(s=-1)}= \bar a^4 \sum_{y_0= 0}^{T'}\sum_{\bf y}\sum_\mu \bar \psi (y) \left\lbrack 
\gamma_\mu \tilde \partial_\mu  +i \frac{\bar a }{2}\gamma_5 \tau_{\mu 5}
 \Delta_\mu   \right\rbrack \psi (y).
\ee
The usual SF boundary conditions can be recovered by performing a chiral rotation of the fermionic 
fields,
\be
\psi'(y) = R(\alpha) \psi(y), \quad \bar \psi'(y) = \bar \psi(y) R(\alpha),
 \qquad R(\alpha) = \exp (i \textrm{$\frac \alpha 2$}  \gamma_5 \tau_{05}).
\ee
For $\alpha =\frac \pi2$ the boundary conditions become the usual ones~(\ref{SFBC}), due to
$R(\textrm{$\frac \pi 2$})Q_{\pm}R^{-1}(\textrm{$\frac \pi 2$}) = P_\pm$.
%
%\be
%\rho(y) = R(\textrm{$\frac \pi 2$})\hat \rho ({\bf y}), \qquad \bar \rho = \hat{\bar \rho}({\bf y})R %(\textrm{$\frac \pi 2$}),
%\ee
%
%and analogously for $\rho'({\bf y}), \bar \rho'({\bf y})$. 
For homogeneous boundary conditions, the action in the standard SF basis takes the form, 
\be
S^{(s=-1)}_{SQ}= \bar a ^4 \sum_{y=0}^{T'}\sum_{{\bf y}} \bar \psi'(y)\left\lbrack\sum_{k} 
\gamma_k \mathcal D_k + \gamma_0 \tilde \partial _0 + \frac {\bar a}{2}\Delta_0 \right \rbrack \psi '(y),
\ee
with ${\mathcal D}_k = \tilde \partial _k +i \textrm{$\frac {\bar a }{2}$}
\gamma_k \gamma_5\tau_{k 5}\Delta_k$.
%
%It is important to remark here that the fields $P_- \psi'(0,{\bf y})$ and $P_+ \psi'(T, {\bf y})$ 
%are now dynamical fields, in contrast with the situation for Wilson quarks. 
%
\subsection{Case $T' = T +a$}
\begin{figure}[ht!]
\begin{center}
\includegraphics{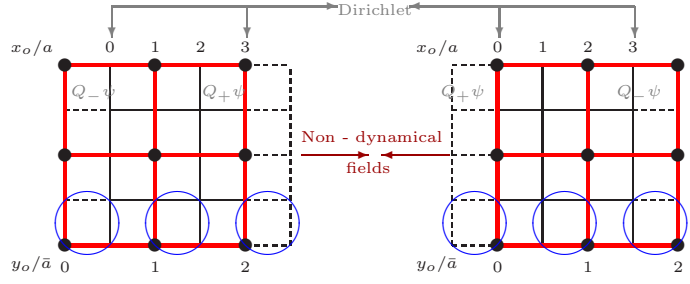}
\vspace{-3mm}
\caption{\label{figura2} Reconstruction of the spinors on a $T = L - a$ lattice. Left $s= 1^+$, right, $s = 1 ^-$}
\end{center}
\vspace{-7mm}
\end{figure}

Here, we distinguish two alternative ways of reconstructing the fermions,
as illustrated in Figure \ref{figura2}. We have labelled the two reconstructions 
with $s = 1^\pm$, according to the sign in front of $\xi_0$ in Eq.~(\ref{reconstr}).
%
%%%%%%%%%%%%%%%%%%%%%%%%%%%%%%%%%%%%%%%%%%%%%%%%%%%%%%%%%%%%%%%%%%%%%
%%%%%%%%%%%%%%%%%%%%%%%%%%%FIGURE%%%%%%%%%%%%%%%%%%%%%%%%%%%%%%%%%%%%
%%%%%%%%%%%%%%%%%%%%%%%%%%%%%%%%%%%%%%%%%%%%%%%%%%%%%%%%%%%%%%%%%%%%%
%%%%%%%%%%%%%%%%%%%%%%%%%%%%%%%%%%%%%%%%%%%%%%%%%%%%%%%%%%%%%%%%%%%%%
%%%%%%%%%%%%%%%%%%%%%%%%%%%%%%%%%%%%%%%%%%%%%%%%%%%%%%%%%%%%%%%%%%%%%
%%%%%%%%%%%%%%%%%%%%%%%%%%%%%%%%%%%%%%%%%%%%%%%%%%%%%%%%%%%%%%%%%%%%%
%
\ba\label{reconstr}
s= 1^+\qquad \qquad &\qquad& \qquad \quad\qquad s = 1 ^ -\nonumber \\
x_0 = 2y_0-a+ a\xi_0, \quad {\bf x} = 2 {\bf y} + a{\boldsymbol \xi}, 
&\qquad & x_0 = 2y_0 - a\xi_0,\quad {\bf x} = 2{\bf y}+ a{\boldsymbol \xi},\nonumber\\
\psi_{\alpha a}(y) = \textrm{$\frac 14$} \sum_{\xi}(\tilde \Gamma_\xi)_{\alpha a }\chi_{\xi}(y),
\qquad\textrm{\hspace{1mm} } &\qquad & \psi_{\alpha a}(y)= \textrm{$\frac 14$}
 \sum _{\xi}(\Gamma_\xi)_{\alpha a } \chi_\xi(y), \\
\bar \psi_{a\alpha}(y)=- \textrm{$\frac 14$}\sum_{\xi}\bar 
\chi_\xi(y)(\tilde \Gamma_\xi^\dagger)_{a \alpha}
,\quad\textrm{\hspace{3mm}}&\qquad&\bar \psi_{a \alpha}(y) = \textrm{$\frac 14$}
\sum_{\xi}\bar \chi_\xi(y)(\Gamma_\xi^\dagger)_{a\alpha }\nonumber \\
\tilde \Gamma_{\xi} = \textrm{$\frac 12$} (-1)^{\xi_0}\gamma_0^{\xi_0},
 \gamma_1^{\xi_1}\gamma_2^{\xi_2}\gamma_3^{\xi_3}.\hspace{1mm}\quad\quad  &&\nonumber 
\ea
The interpretation of Figure~\ref{figura2} is the same as Figure~\ref{figura1}. 
Eqs.~(\ref{reconstr}) specify how to reconstruct the 
four-component fermions in both cases. Depending on the reconstruction, the boundary conditions
are different, and therefore the chiral rotations needed to restore the usual 
SF b.c.'s differ, too:
\ba
s= 1^+ \qquad\qquad\qquad\qquad\quad & \qquad & \qquad\qquad \qquad  s= 1^- \nonumber \\
Q_-\psi(0,{\bf y})= \hat \rho,\hspace{3mm}Q_+ \psi(0,{\bf y})=\hat \rho',\hspace{2mm}\quad\qquad&\qquad & 
Q_+\psi(0,{\bf y})= \hat \rho,\hspace{3mm}Q_- \psi(0,{\bf y})=\hat \rho', \nonumber \\
\bar \psi(0,{\bf y})Q_- = \hat{\bar\rho},\hspace{3mm}
\bar \psi'(0,{\bf y})Q_+ = \hat{\bar \rho}',\hspace{2mm}\quad\qquad&\qquad & 
\bar \psi(0,{\bf y})Q_+ = \hat{\bar\rho}\hspace{3mm}
 \bar \psi'(0,{\bf y})    Q_- = \hat{\bar \rho}', \\
\psi'(y) = R({\scriptstyle -\frac\pi2})\psi(y),\quad 
\bar \psi'(y) = \bar \psi(y)R({\scriptstyle - \frac \pi 2}), &\qquad &
\psi'(y) = R({\scriptstyle\frac \pi2})\psi(y),\quad\bar
 \psi'(y) = \bar \psi(y)R({\scriptstyle\frac\pi2}) . \nonumber
\ea
However, once rotated to the standard SF basis, the action for both cases $s = 1^\pm$ is the same,
\be
S^{(s=1)}_{SQ} = \bar a ^4 \sum_{y_0,{\bf y}}\bar \psi'(y)\left\lbrack\sum_{k}\gamma_k \mathcal D_ k
 + \gamma_0 \tilde \partial_0 - \frac{\bar a}{2}\Delta_0 \right \rbrack\psi'(y).
\ee
\subsection{Symmetries of the SF with staggered fermions}
The symmetries of the SF with staggered quarks have been summarised by Heller~\cite{Heller:1997pn}. 
We here just wish to emphasise
that the flavour and chiral symmetries refer to a particular basis. As we have seen, the boundary conditions may depend on the way the four-spinors are reconstructed. It is only after performing a chiral non-singlet rotation that the standard SF is recovered. In this basis, the usual axial U(1) symmetry of staggered quarks becomes a flavour symmetry. More precisely, the transformation
%
%In the standard staggered basis, there is a chiral symmetry left, that reads like this,
%
\be\label{chiral}
\psi(y) \rightarrow e^{i\beta\gamma_5\tau_5}\psi(y), \qquad\bar \psi(y) \rightarrow \bar \psi(y)e^{i\beta\gamma_5\tau_5},
\ee
when rotated into the SF basis, becomes a continuous flavour symmetry with generator $\tau_0$, 
\be
\psi'(y) \rightarrow e^{i \beta \tau_0}\psi'(y),\qquad \bar \psi'(y) \rightarrow \bar \psi'(y) e^{-i \beta \tau_0}.
\ee
Furthermore we note that spatial translations by a unit $a$ on the fine lattice.
\be\label{flavor}
\psi(y)\rightarrow\tau_k \psi(y) + \bar a \tau_kQ_{+}{(k)}\partial_k \psi(y)\qquad
\bar \psi(y) \rightarrow \bar \psi(y)\tau_k + \bar a \bar \psi(y){ }^{{ }^{{ }^\leftarrow}}
{\hspace{-2.5mm}}\partial_k\tau_k Q_+^{(k)},
\ee 
with $Q^{(k)}_\pm = \frac 12 (1 \pm \gamma_k\gamma_5\tau_{k5})$,
correspond to a discrete subgroup of flavour symmetry in the continuum limit.
%
%%%%%%%%%%%%%%%%%%%%%%%%%%%%%%%%%%%%%%%%%%%%%%%%%%%%%%%%%%%%%%%%%%
%%%%%%%%%%%%%%%%%%%%%%%%NEW SECTION%%%%%%%%%%%%%%%%%%%%%%%%%%%%%%%
%%%%%%%%%%%%%%%%%%%%%%%%%%%%%%%%%%%%%%%%%%%%%%%%%%%%%%%%%%%%%%%%%%
\vspace{-4mm}
\section{Correlation functions at tree level}
To evaluate the correlation functions we first integrate over the quark fields.
The expectation value assumes the form  $\langle \mathcal O\rangle = 
\langle[\mathcal  O]_F \rangle_G $, where $\langle\rangle_G$  denotes the gauge field average.
We have determined the free quark propagator both analytically and numerically and may therefore
compute the correlation functions to tree level.
The remaining chiral symmetry of Eq.~(\ref{chiral}) becoming a flavour symmetry with generator $\tau_0$ 
in the standard SF basis, disconnected diagrams in the computation of 
$f_A, f_P, f_1$ are forbidden if we choose those flavour matrices which anticommute with $\tau_0$. 
For these matrices, the correlation functions $f^{ab}_A(y_0)$ reads,
\be
f^{ab}_{\rm A}(y_0) =\bar a^6\sum_{\bf y', y''}
\frac18 \left\langle {\rm tr}\left(\lbrack\zeta({\bf y''})\bar \psi'(y) 
\rbrack_F\gamma_0\gamma_5 \tau^a\lbrack\psi'(y)\bar \zeta({\bf y'}) 
\rbrack_F\gamma_5     \tau^b\right) \right\rangle_G, 
\ee
and analogous expressions are obtained in the other cases.
The continuum values of $f_{\rm X}^{ab}$ at tree level with vanishing background field
take the form, $\delta^{ab}f_{\rm X}$, with
\be
f_{\rm A}(T'/2) = -\frac{N_c}{\cosh^2(\sqrt 3 \theta)}, 
\quad f_{\rm P}(T'/2) = \frac{N_c}{\cosh(\sqrt 3 \theta)},
\quad f_1 = \frac{N_c}{\cosh^2(\sqrt 3 \theta)}.\nonumber
\ee
where $\theta$ is a phase factor coming from the generalised 
boundary conditions, i.e. $\psi(y + L\hat k) = e^{i\theta}\psi(y), 
\bar \psi(y + L \hat k) = \bar \psi(y)e^{-i\theta}.$ Including the correct tree level 
boundary  counterterm, the results obtained  are accurate up to 
${\rm O }(a^2)$ for $f_{\rm P}, f_1$ and ${\rm O}(a)$  for $f_{\rm A}$. 
%
%%%%%%%%%%%%%%%%%%%%%%%%%%%%%%%%%%%%%%%%%%%%%%%%%%%%%%%%%%%%%%%%%%
%%%%%%%%%%%%%%%%%%%%%%%%NEW SECTION%%%%%%%%%%%%%%%%%%%%%%%%%%%%%%%
%%%%%%%%%%%%%%%%%%%%%%%%%%%%%%%%%%%%%%%%%%%%%%%%%%%%%%%%%%%%%%%%%%
\vspace{-3mm}
\section{Fermionic O($a$) improvement}
\vspace{-2mm}
\subsection{Infinite volume}
Close to the continuum limit, the lattice theory may be described in terms 
of a local effective theory with action \cite{Symanzik:1983dc},  
\be
S_{\rm eff} = S_0 + a S_1 + a^2S_2+ \dots ,\qquad 
S_k = \int d^4y \mathcal L_k(y) 
\ee
The apparent ${\rm O}(a)$ contributions on the infinite lattice
are fixed by the shift symmetry, since it is the combined expression 
$\mathcal D_\mu \gamma_\mu$ that is invariant under this 
transformation and not the usual kinetic term alone 
As was pointed out in \cite{Luo:1996vt}, there are no invariant dimension 5 operators,
so no counterterms  can be added. The standard procedure to eliminate the apparent O($a$) terms
consists in defining improved field,
\ba
\psi^I(y) &=& \psi(y) +\frac{ \bar a}{4}\sum_{\nu}  
(Q^{(\nu)}_+ - Q^{(\nu)}_-) \tilde \partial_\nu\psi(y),\nonumber\\
\bar \psi^I(y) &=& \bar \psi(y) + \frac{\bar a}{4} \sum_\nu 
\bar \psi \Dizq {\textrm{\footnotesize$\tilde \partial_\nu$}}(y)
  (Q^{(\nu)}_+- Q^{(\nu)}_-),
\ea
in terms of which one finds,
\be
S_{SQ} = \bar a^4 \sum_{y\mu}\bar \psi^I(y)\gamma_\mu \tilde
 \partial_\mu \psi^{I}(y) + {\rm O}(a^2).
\ee
\vspace{-8mm}
\subsection{O($a$) effects from the boundaries}
In the SF framework, additional renormalisations and O($a$) cutoff effects may arise 
from the very presence of the boundaries. Taking the symmetries into account, 
we arrive at the conclusion that there is only one possible  dimension 3 operator, $\bar \psi'\psi'$. This is the same as
encountered for Wilson quarks, and can thus be absorbed in a multiplicative renormalisation
of the quark and anti-quark fields at the boundaries. 

In the case of dimension 4 operators, we obtain again the same result as for Wilson quarks. 
However, when using the equations of motion, we here prefer a different choice for the
counterterm action, namely
\be
\delta S_{F,b}[U, \bar \psi, \psi] = \bar a^4 \sum_{\bf y}\left\{
(d_{1} - 1)\lbrack\hat {\mathcal O}_{b,1} +
\hat{ \mathcal O}'_{b,1} \rbrack + (d_{2} - 1)\lbrack 
\hat{\mathcal O}_{b,2} + \hat{\mathcal O}'_{b,2} \rbrack \right\},
\ee
\vspace{-5mm}
\ba
\hat{\mathcal O}_{b,1} = \bar \psi' (0,{\bf y})P_+\gamma_k\mathcal 
D_k \psi'(0,{\bf y}),\hspace{1mm} && \hat{\mathcal O}_{b,2} = 
\bar \rho({\bf y})\gamma_k \mathcal D_k \rho({\bf y}), \\
\hat{\mathcal O}'_{b,1} = \bar \psi (T,{\bf y})P_-\gamma_k\mathcal 
D_k \psi(T,{\bf y}),\hspace{2mm} && \hat{\mathcal O}'_{b,2} = 
\bar \rho'({\bf y})\gamma_k \mathcal D_k \rho'({\bf y}), 
\ea
The coefficients $d_{1,2}$ have a perturbation
expansion in powers of $g_0^2$. We have determined $d_{1}$ at tree 
level,
\be
 d_1^{(0)} \Big|_{T' = T\pm a}\hspace{-2mm}= 1 \pm \frac 14.
\ee
\vspace{-5mm}
%
%%%%%%%%%%%%%%%%%%%%%%%%%%%%%%%%%%%%%%%%%%%%%%%%%%%%%%%%%%%%%%%%%%%
%%%%%%%%%%%%%%%%%%%%%%%%NEW SECTION%%%%%%%%%%%%%%%%%%%%%%%%%%%%%%%
%%%%%%%%%%%%%%%%%%%%%%%%%%%%%%%%%%%%%%%%%%%%%%%%%%%%%%%%%%%%%%%%%%
\vspace{-5mm}
\section{Conclusions}
We have reconstructed the four-component spinors in the SF with staggered quarks, for both
cases $T' = T \pm a$, computed the free propagator and the correlation functions 
$f_{\rm A}, f_{\rm P}, f_1$ at tree level.
The implementation of ${\rm O}(a)$ improvement  is work in progress. 
Once it is fully understood, we plan to trace the running of the SF 
coupling and the quark mass in four-flavour QCD
%
%
%%%%%%%%%%%%%%%%%%%%%%%%%%%%%%%%%%%%%%%%%%%%%%%%%%%%%%%%%%%%%%%%%%%
%%%%%%%%%%%%%%%%%%%%%%%%NEW SECTION%%%%%%%%%%%%%%%%%%%%%%%%%%%%%%%
%%%%%%%%%%%%%%%%%%%%%%%%%%%%%%%%%%%%%%%%%%%%%%%%%%%%%%%%%%%%%%%%%%%

\end{document}